\documentclass[amsmath,amssymb,twocolumn,showpacs,prl]{revtex4}
\usepackage{amsmath,graphicx,dcolumn,bm}

\begin{document}
\title{Random particle packing with large particle size variations using reduced-dimension algorithms}

\author{M. D. Webb}
\email{michael.webb@atk.com}
\author{I. Lee Davis}
\email{lee.davis@atk.com} \affiliation{ATK Thiokol\\P. O. Box 707
\\Brigham City, Utah 84302}

\date{\today}

\begin{abstract} We present a reduced-dimension, ballistic
deposition, Monte Carlo particle packing algorithm and discuss its
application to the analysis of the microstructure of hard-sphere
systems with broad particle size distributions. We extend our
earlier approach (the ``central string'' algorithm) to a
reduced-dimension, quasi-3D approach. Our results for monomodal
hard-sphere packs exhibit a calculated packing fraction that is
slightly less than the generally accepted value for a maximally
random jammed state. The pair distribution functions obtained from
simulations of composite structures with large particle size
differences demonstrate that the algorithm provides information
heretofore not attainable with existing simulation methods, and
yields detailed understanding of the microstructure of these
composite systems.
\end{abstract}

\maketitle

\section{Introduction}

Hard-sphere particle packing models have proven useful in the study
of liquids and fluids \cite{bernal59,bernal60,scott69,martys94},
glasses \cite{truskett00,kim03}, foams \cite{ohern01}, granular
flows \cite{kadanoff99,makse02}, and amorphous solids
\cite{bennett72,srolovitz81}, and have been extended to
non-spherical particles as well \cite{coelho97,donev04a}. Perhaps
the most interesting application of particle packing is its use as a
tool to understand the microstructure of particulate materials and
powders. Understanding order in particulate systems is an
outstanding question, and particle packing approaches have
contributed to this line of research for some time
\cite{bernal59,bernal60,bennett72,berryman83,jullien96,quintanilla96,rintoul96a,rintoul96b,kansal00,truskett00,torquato01,kansal02}.
Many researchers also use the tool to study a variety of related
lines of research, including the interpretation of amorphicity as
spatial chaos in one dimension \cite{Reichert84}, pressure and
entropy in crystals \cite{speedy98}, thermodynamics of slowly
sheared granular systems \cite{makse02}, the onset of dilatancy in
loose packings \cite{onoda90}, transport properties
\cite{truskett00}, and porosity \cite{yang96,yang00,truskett00} to
name a few. Of particular interest is the recent work on the concept
of random close packing and the more rigorous definition of jamming
in these systems \cite{torquato00,kansal02,ohern03,donev04b}.

Particle packing simulations take a variety of forms. Early attempts
incorporated real particles such as powders \cite{mcgeary61}, ball
bearings \cite{scott64,scott69,finney70}, balls and spokes
\cite{bernal59,bernal60}, and recently glass beads in a neutrally
buoyant fluid \cite{onoda90}, horizontally shaken beads
\cite{pouliquen97}, and M\&M\footnote{M\&M's Candies is a registered
trademark of Mars, Inc.} chocolate candies \cite{donev04a}.
Numerical and Monte Carlo techniques include seed aggregation
\cite{bennett72} or variants thereof \cite{matheson74}, growth in a
unidirectional force field \cite{visscher72}, shrinking of randomly
placed spheres \cite{Jodrey85}, dynamic, growing spheres on the
surface of a hypersphere \cite{Tobochnik88}, overlap relaxation
followed by space expansion and vibration \cite{he99}, conjugate
gradient energy minimization \cite{ohern02,ohern03}, and various
forms of molecular dynamics
\cite{lubachevsky90,lubachevsky91,elliot00,kim03}.

\section{Motivation for this work}
We seek to calculate the bulk and microstructural properties of
particulate systems composed of rigid or semi-rigid particles
embedded in an elastic or viscoelastic matrix
\cite{davis93,davis99cosm}. This area of research demonstrates that
the macroscopic properties of a composite material (e.g., the
modulus) can depend on the microscopic details of the packing
structure comprising the material \cite{hubner00,hatchpc}. For
example, the bulk response of a particulate system to an applied
external stress can be determined in large part by particles in the
pack that lie very near each other (i.e., within a fraction of a
particle radius). For this reason, our desire to calculate bulk
properties using a first-principles approach relies on a thorough
understanding of the material microstructure. Moreover, detailed
knowledge of the microstructure of such systems allows for related
studies, such as combustion of solid rocket propellants
\cite{kochevets01}.

For some systems of interest, the particulate ingredients that
comprise the system may include particles with potentially very
large size differences. Packing simulation of such systems with
traditional methods can be problematic as the size difference
increases. For example, a simulation of spherical particles with a
size ratio of 100:1 would require $10^6$ small spheres for each
large sphere in the pack (assuming an equal mass ratio). Obtaining
good statistics for such packs would require literally billions of
particles in the pack, which is clearly unattainable with extant
computational methods.

In an earlier paper, one of us introduced the concept of the central
string algorithm \cite{davis90}. The key feature of that approach is
the idea that some packing statistics near a line drawn through a
composite structure ought to mimic the three-dimensional statistics
of the pack for a sufficiently long line. We emphasize that a
reduced-dimension algorithm may not be able to supplant a 3D
algorithm for all statistics of interest. In this paper, we focus on
particle packing fraction and radial distribution functions; the
treatment of other statistics is under investigation.

The appeal of the reduced-dimension approach is that, when correctly
implemented, it should allow researchers to calculate certain
statistics for packs with large particle size variations which would
otherwise be unattainable due to the long computation times
associated with a full three-dimensional packing of such composite
materials.

The earlier work \cite{davis90} demonstrated the promise of a
reduced-dimension approach, but the central-string method exhibited
shortcomings. The most serious shortcoming present in the previous
model was that the perturbation approach was unable to prevent
particle segregation during growth. Also, the earlier model was
unable to provide detailed information about the microstructure
because of the limited structure found around the central string.
The present work seeks to overcome these shortcomings by extending
the central-string approach to a quasi-3D method, thereby providing
the ability to calculate detailed microstructure, including radial
distribution functions and coordination numbers.

Our algorithm follows most closely that of Visscher and Bolsterli
\cite{visscher72}, and is in essence a ballistic dispersion,
reduced-dimension Monte Carlo simulation.

\section{The concentric-cylinder, reduced-dimension approach}

A natural extension of the central-string approach to
reduced-dimension particle packing is to extend the central string
into a cylinder, producing a quasi-3D packing algorithm. In effect,
we extend the notion of a line drawn through the three-dimensional
particle pack to a cylinder drawn through the particle pack. Whereas
the particles that intersected the line represented the packing
statistics (for a sufficiently long line), a cylinder of equivalent
length which cuts through the pack should also represent the pack
statistics, but with more accuracy (per unit length) because more
particles are included in the cylinder than intersect the line. In
particular, if properly simulated and interpreted, this approach
should provide sufficient information to allow a detailed
examination of the pack microstructure.

\begin{figure}
\includegraphics[scale=0.40]{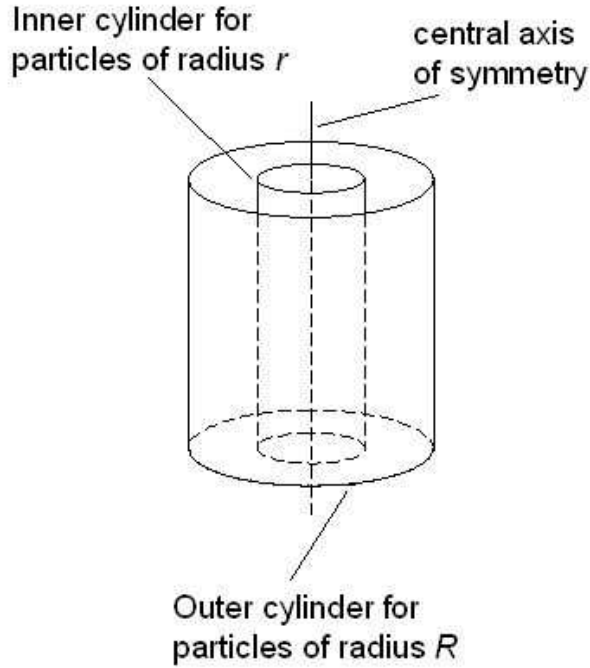}
\caption{\label{fig:concencyl} Representation of the reduced-dimension, concentric-cylinders geometry.}
\end{figure}

The key to this reduced-dimension approach is to represent each
particle of a different size or density (a {\em mode}) in the pack
by its own cylinder, scaled appropriately to the particle's size. We
define a particle mode to be all particles in a pack that are
indistinguishable relative to one or more properties of interest. A
monomodal pack comprises a single cylinder (and is in fact
equivalent to a three-dimensional pack). A binary pack includes two
concentric cylinders, whose radii are scaled proportionally to the
size of each particle. A ternary pack comprises three cylinders, and
so on; see Fig. \ref{fig:concencyl}. All cylinders share a common
axis, but the cylinder radii depend on each particle's size.

We construct the pack by dropping particles at a randomly chosen
position within each particle's cylinder. When simulating more than
one particle mode, the order in which the particles are dropped is
also random (respecting the final number of each particle mode
required to represent the desired mass fraction of each mode). Each
particle is dropped above the pack and allowed to descend into the
pack under the influence of a unidirectional force field (e.g.,
gravity) acting along the $z$-axis until the particle finds a
contact stability point. Until the particle finds a stability point,
it will roll along other particles or the roll corridor defined by
one or more particles and/or the cylinder wall. Contact stability
refers to whether the current sphere is in compressive contact with
an object or in tensile contact.  If it is in compressive contact
with another sphere or the cylinder wall, the particles push against
each other due to the current sphere's weight. When a particle
touches three or more objects compressively, it is stable and is
placed at that position. When in tensile contact with another
particle or the cylinder wall, the current sphere will roll away
from the object unless it already has three or more compressive
contacts. After the particle is placed at a contact stability point,
another particle is dropped and the sequence repeated until
completion (all particles dropped and stable).

The dropping and rolling of pack particles does not include any
dynamics.  The particles do not bounce, do not gain speed or shoot
off the edge of another particles with a parabolic trajectory. Once
placed at a stability point, the particle remains there for the
duration of the simulation. It is as if the particles are being
dropped in a highly viscous fluid so that all inertia is absent and
the particle creeps to its final resting place. In general, we use a
number of trial drops (8-32) for each particle and choose the lowest
of the set to ensure the densest packing behavior.

As we build the pack, each particle mode remains within its own
cylinder. The smallest mode particle resides in the smallest
cylinder, and is closest to the axis of symmetry. Larger particles
may lie within the inner cylinders, but also extend into outer
cylinders; each mode is always constrained to remain within its own
cylinder. The largest particle in the pack may lodge in any of the
cylinders, and in the outermost cylinder it resides alone. It should
be clear that setting all cylinders to be the same size (the size of
the largest particle's cylinder) produces a three-dimensional
simulation. The implementation we have chosen thus allows us to
build simulated packs with both reduced- and three-dimensional
approaches by simply choosing each particle's cylinder size
appropriately.

\subsection{The power of the reduced dimension approach}
The development of the reduced-dimension approach to particle
packing provides the opportunity to analyze particle packs with
broad size distributions which would otherwise be computationally
prohibitive. A full, three-dimensional simulation of 600,000
particles in a ternary pack with size ratios of 30:50:175 consumes
40 hours (real-time) on a fast desktop PC, placing approximately
four particles per second. A reduced-dimension simulation which
produces comparable statistical results runs in two minutes, placing
170 particles per second. This computational advantage increases
tremendously as the particle size ratio increases. We recently
investigated a composite material consisting of four particles of
varying sizes with a maximum size ratio of 77:1; the
reduced-dimension simulation modeled $3\cdot10^6$ particles and
required about 100 hours to complete. We estimate that a
three-dimensional simulation of the same fidelity would have
required about $15\cdot10^{9}$ particles and about 100 years
computer time with our simulation.

In general, we find that for equal numbers of particles, the
algorithm's time to completion as a function of the the particle
size ratio $R$ varies as $R^{3.3}$ for a fully three-dimensional
simulation, but only as $R^{0.4}$ using the reduced dimension
algorithms. Using the reduced-dimension approach, the algorithm time
to completion as a function of the number of particles $N$ is linear
in $N$, with a particle size ratio of about 7-10:1. In three
dimensions, the same calculation varies as $N^{3}$. The
reduced-dimension algorithm brings the simulation of complex
particle packs with large particle size variation into the realm of
possibility.

\section{Implementation of the concentric cylinders algorithm}
The introduction of concentric cylinders of arbitrary size into the
basic ballistic deposition algorithm introduces challenging
analytical and simulation issues. The inclusion of the
arbitrary-size cylinders must not be allowed to change the
microstructure and must not introduce artificial structures into the
simulation.

\subsection{Particle segregation}
The primary problem introduced by the concentric-cylinder,
reduced-dimension algorithm is particle segregation during pack
growth. Particle segregation takes one of two forms: segregation in
the growth direction (axial segregation), and segregation in the
radial direction (normal to the growth direction). Segregation in
the growth direction may occur for any particle in the simulation,
while segregation in the radial direction may occur for all but the
smallest particle in the simulation.

\subsubsection{Axial segregation}
Segregation in the growth direction for small particles is a
``real'' phenomenon for particle packing in a unidirectional force
field. In essence, the small particles fall through the interstitial
spaces between bigger particles until they find a resting place. The
final resting position can be far below the current pack ``surface''
if the pack is porous (as compared to the small particle's radius).
This effect is an artifact of the simulation method, and doesn't
represent real composite materials that are prepared through
adequate mixing or in the presence of an interstitial matrix or
binder.

To prevent axial segregation of small particles, we introduce the
concept of the {\em catch net}. The catch net prevents particles
from falling too far (typically a few radii) below the growing pack
surface, even if there is an allowed pathway. The catch net position
after the addition of the $i$-th particle, $h_{cn}^{i}$ is:

\begin{equation}
\label{eq:cnpos} h_{cn}^{i}=A_z\left(z_i -
r_j,N_{cn}\right)-2r_j\left(1+u\right)
\end{equation} where $z_i$ is the height (position in the growth
direction) of the $i$-th particle, $r_j$ is the radius of the
particle (of mode $j$), $u$ is a uniform random deviate on the range
(0,1), $N_{cn}$ is the length of the moving average, as given by

\begin{equation}
\label{eq:ncn} N_{cn}=\frac{R_{0}^{2}}{r_{0}^{2}},
\end{equation} with $R_0$ equal to the radius of the confining
cylinder of the smallest mode, and $A_z(x,N)$ is the exponential
moving average of length $N$ of the time-series variable $x$:

\begin{equation}
\label{eq:ema}
A_z\left(x_i,N\right)=\frac{2}{N+1}x_i+\frac{N-1}{N+1}A_z\left(x_{i-1},N\right)
\end{equation} where

\begin{equation}
\label{eq:ema0} A_z\left(x_1,N\right)=x_1.
\end{equation} In the current implementation, the default value
chosen for $N_{cn}$ is the number of particles required to ``tile''
the innermost cylinder, as given by Eq. (\ref{eq:ncn}). Particles
that would otherwise fall below the catch net surface are prevented
from doing so, are fixed at the current position of the catch net,
and remain at that position during the remaining pack growth. The
catch net surface only acts on particles with the smallest radius in
the innermost cylinder.

Particles larger than the smallest particle may also experience
segregation in the growth direction. The segregation of these
particles is a direct artifact of the reduced-dimension algorithm.
When the larger particles fall in an outer cylinder, they do not
have smaller particles to support them, so they fall until they
lodge against other large particles. If the large particles are
densely populated in the pack, segregation is not an issue, as they
will support each other at the appropriate position. If the large
particles are sparsely populated in the pack, a new particle may
fall many diameters below what would otherwise be the pack surface.
This undesired effect is a direct consequence of the fact that the
reduced-dimension algorithm does not include all particle types in
the cylinders outside the innermost cylinder.

To remedy the segregation of these particles in the growth
direction, we introduce the concept of a synthetic surface. The
synthetic surface serves to support particles that would otherwise
fall below the surface defined by the particles in the innermost
cylinder. The definition of the synthetic surface position,
$h_{ss}^{i}$, after the addition of the $i$-th particle is:
\begin{equation}
\label{eq:synsurf}
h_{ss}^{i}=\frac{\sum_{k=1}^{N_{ss}}\left(z_k+r_k\right)}{N_{ss}}+r_i,
\end{equation} where $z_k$ is the height of the $k$-th particle,
$r_k$ is the radius of the $k$-th particle, and $N_{ss}$ is the
number of spheres included in the average. It is important to point
out that the sum is over the $N_{ss}$ particles with the highest
north poles in the pack, not the last $N_{ss}$ particles added. (The
``north pole'' is the position of the top of the sphere--the center
of the sphere plus the sphere radius.) Any particle whose center
would fall below the synthetic surface position is prevented from
doing so, fixed at the current position of the synthetic surface,
and remains at that position during the remaining pack growth. The
synthetic surface is a single surface active for all particles in
the pack except the particles in the innermost cylinder (which is
managed by the catch net). In the current implementation, $N_{ss}$ =
$N_{cn}$, as given by Eq. (\ref{eq:ncn}).

\subsubsection{Radial segregation}

Particle segregation in the radial direction is also an artifact of
the reduced-dimension approach. The effect is most pronounced for
particles whose radius is significantly larger than the radius of an
inner cylinder. These particles tend to roll off the pedestal formed
by particles in the inner cylinder, because uneven variations in the
surface structure act as inclined planes. If the particle radius is
sufficiently large, the moment arm above these planes is big enough
to cause the particle to roll out of the inner cylinders. A
straightforward calculation of the geometry of the situation shows
that rolling can occur for particle radii ratios as small as a few
to one.

To remedy this situation, we force particles to remain within the
largest cylinder that encloses the particle's center position upon
selection of its drop point. Choosing drop points to be uniformly
randomly distributed across all allowed cylinders ensures that each
cylinder contains the correct number of particles, accurately
representing a three-dimensional pack. Examination of the radial
distribution of each particle mode in the resulting pack shows that
this approach accurately maintains the expected $r^2$ dependence of
the particle distribution.

\subsection{Edge effects}

Another potential problem that can be important but is easily
negated is an edge effect, which occurs when the radius of the
confining cylinders is not large compared to the particle radius. If
the confining cylinders for each particle mode are chosen to be not
much bigger than the particle radius (e.g., just a few times larger
than the particle radius), then edge effects tend to dominate the
particle statistics near the cylinder walls. This undesired effect
can be substantial: for some configurations, these edge effects
reduce the population of particles near the cylinder walls by as
much as 50 percent from the nominal values; the particles are no
longer distributed as $r^2$. Fortunately, this effect has a simple
remedy; we set the confining cylinder for each particle to be at
least 20 times the particle radius . Tests show that this reduces
the magnitude of the variation from the nominal value in particle
population at the edges to less than one percent. This causes the
simulation to run slower than it would with a smaller cylinder, but
is required in order to ensure pack integrity.

\subsection{Implementation summary}
In summary, we introduce a reduced-dimension approach to the
standard ballistic deposition method. Concentric cylinders contain
particles of various sizes, with increasing cylinder diameter
corresponding to increasing particle size. The particles are
distributed as $r^2$ within each cylinder, out to each particle's
outermost constraining cylinder. The presence of small particles is
simulated in the outermost cylinders (where they do not reside) by
calculating the effective surface height of these particles and
representing that surface throughout all larger cylinders. We
prevent small particles from percolating downward through large
interstitial voids in the pack. We consider the algorithm to be
``quasi-3D'' because each cylinder is at least 20 times the radius
of the particle it confines.

Detailed analysis of the simulation results, including the resulting
pack structure, demonstrates conclusively that the reduced-dimension
approach produces numerically identical results (within statistical
uncertainty) to three-dimensional simulations, but at much less
cost. In some cases, the reduced-dimension algorithm allows
calculation of microstructure for packs that would otherwise be
unattainable with existing methods.

Each of the results presented in this paper has been verified by
direct comparison to three-dimensional simulations which do not
include any of the reduced-dimension algorithms. In the discussion
to follow, we present comparison results between reduced-dimension
and 3D simulations that verify this claim.

\begin{figure}
\includegraphics[scale=0.5]{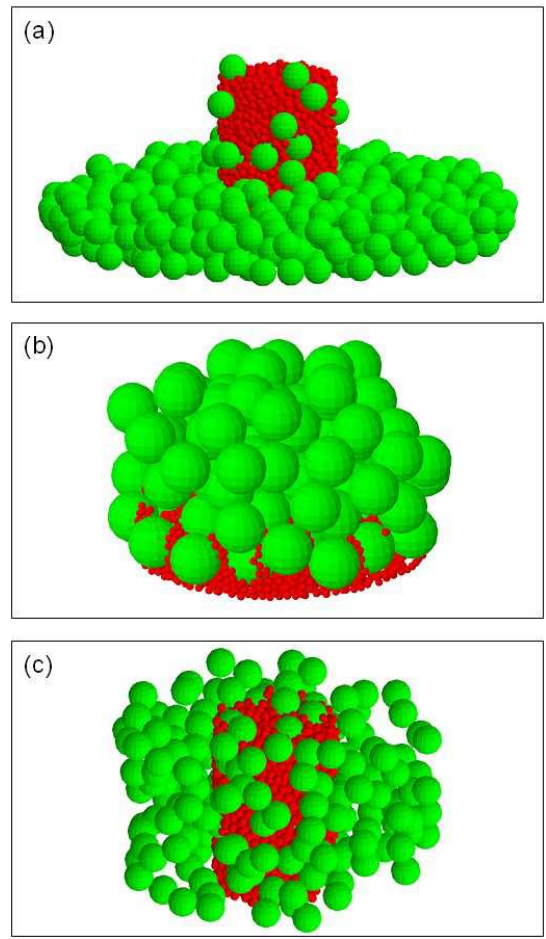}
\caption{\label{fig:packs}(a) The effects of large particle
segregation, before implementation of the synthetic surface. (b) The
effects of small particle segregation, before implementation of the
catch net. (c) A representative pack with all reduced-dimension
algorithms implemented; much of the outermost cylinder has been cut
away in this view so that the innermost cylinder is not masked.}
\end{figure}

In figure \ref{fig:packs}, we show representations of particle packs
obtained from the simulations. Inset (a) shows the effects of axial
segregation of large particles when the synthetic surface is not
implemented. Inset (b) shows the effects of axial segregation of
small particles when the catch net is not implemented. Inset (c)
shows a reduced-dimension pack with all reduced-dimension algorithms
implemented.

\subsection{Limits of the reduced-dimension algorithm}

We haven't performed an exhaustive comparison of results obtained
from 3D and reduced-dimension simulations due to the very long
run-times associated with 3D simulations of particle packs with very
large (e.g., $\geq$10:1) particle size differences. We have
determined limits on some simulation parameters such as cylinder
size, total particle number, and largest-smallest size ratio. As
described above, to avoid edge effects, each cylinder size must be
at least 20 times the particle radius; smaller cylinders yield
unacceptable deviations from uniformity in particle density across
the cylinder. To acquire acceptable statistics, each particle mode
should occur at least $\sim1000$ times in the pack. This operational
limit leads to constraints on the largest-smallest particle size
ratio that can be examined with the current algorithm, and depends
on the computing resources available. Currently, total particle
numbers in the pack should be constrained to less than about
$10\cdot10^6$ particles, meaning that particle size ratios of about
1000:1 are attainable for binary packs, depending on the relative
mass fraction in each mode. We have successfully simulated particle
packs with about $5\cdot10^6$ particles, consisting of about 30
particle modes, with the largest particle radius ratio around 80:1.

\section{Calculation of pack microstructure and statistics}
The concentric cylinders approach introduces complexity into the
calculation of pack microstructure and packing statistics. The
reduced-dimension nature of the pack destroys the three-dimensional
symmetry of the pack; the pack is highly anisotropic. The
calculation of packing fraction and radial distribution functions is
complicated by this structure. We cannot directly calculate
microstructural statistical properties in the usual way for a 3D
pack.

Fortunately, this complication is fairly easy to overcome by
correctly accounting for the asymmetry. Some properties require that
we include only the innermost cylinder's structure in our
calculations, while others require us to renormalize the statistic
to account for the anisotropy in the reduced-dimension pack.

The calculation of volume (packing) fraction is done for only the
innermost cylinder. Each particle's contribution to the packing
fraction simply becomes the ratio of the volume each particle
occupies in the innermost cylinder to the total volume of that
cylinder. (Only the portion of the particle that lies within the
innermost cylinder is included in the volume calculation.)

The calculation of the radial distribution function requires an
accounting of the pack asymmetry. The radial distribution function
$G_{ij}(r)$ is defined to be the number density of particles of type
$j$ that are within the range ring $r$ and $r+dr$ from particles of
type $i$:

\begin{equation}
\label{eq:gijdef} G_{ij}(r)=\frac{n_{ij}}{4\pi r^2}
\end{equation} where $r$ is the range from particle $i$ to particle
$j$, and $n_{ij}$ is the number of particles at the given range
\cite{cusack87,mason68}.

To calculate the radial distribution function for a
reduced-dimension pack, we calculate $G_{ij}$ according to equation
(\ref{eq:gijdef}) within cylinder $i$, and {\em renormalize} the
resulting value by multiplying the total volume of the range ring
within cylinder $i$ by the appropriate value to obtain a full $4\pi$
steradians at the given range. In essence, the contribution to
$G_{ij}$ within cylinder $i$ is weighted according to the fraction
of $4\pi r^2$ contained within cylinder $i$.

We show below that detailed comparisons of three-dimensional results
obtained without any reduced-dimension algorithms to this
calculational approach for the reduced-dimension algorithm confirm
the validity of this approach.

\section{Applications}

\subsection{Comparison to mechanical packing experiments}

We have used this new reduced-dimension algorithm extensively in the
analysis of monomodal and multimodal packs. This allows us to
compare the simulation results with previously published
experimental and simulation data. It's not entirely clear what would
be the best basis for comparison. Mechanical shaking experiments
such as those performed by McGeary\cite{mcgeary61} appear to be the
most applicable experimental work. However, there is known variation
in the results of such experiments; Scott showed that, depending on
the physical methods used, the packing fraction of equal-sized
spheres could vary between 0.60-0.637\cite{scott60}.

A similar, but more dramatic, variation exists in numerical and
simulation methods. In crystalline structures, the simple cubic
structure has a packing fraction of about 0.52, while the
close-packed face-centered cubic lattice and its stacking variants
has a packing fraction of about 0.74. So-called random-close packed
structures generally fall in between these limits, but the very
definition of a random close-packed structure is in
question\cite{torquato00}.

In general, as one would expect, the ballistic deposition approach
used here yields packing fractions that are lower than those
obtained by other methods. For example, the measured monomodal
packing fraction in our simulation is 0.60, about three-four percent
less than that expected for a maximally random jammed pack of equal
size hard spheres \cite{kansal02}.

\begin{figure}
\includegraphics[scale=0.4]{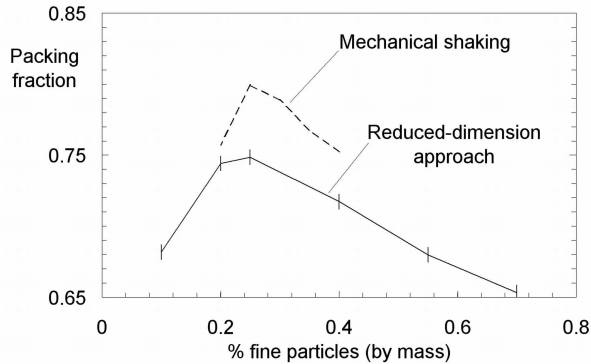}
\caption{\label{fig:comp6-1} Comparison of simulation results to
mechanical packing data \cite{mcgeary61} for a particle size ratio
of 124:19 (about 6.5:1). Error bars on the simulation data are +/-
one standard deviation in all figures.}
\end{figure}

We have compared the packing fractions obtained from our simulation
with those of previously published mechanical shaking experiments,
and find a similar lower value than published measurements for
binary mixtures \cite{mcgeary61}. In Fig. \ref{fig:comp6-1}, we show
a comparison of the packing fractions obtained from \cite{mcgeary61}
and our simulation for a binary pack with size ratio of 124:19
(about 6.5). The data shows close agreement between the packing
fractions at small and large mass ratios, but more deviation for
mass fractions that are approximately equal. However, the
qualitative agreement between the curves is excellent, with the
maximum value of the packing fraction occurring at the same mass
fraction value (25\% fine).

\begin{figure}
\includegraphics[scale=0.40]{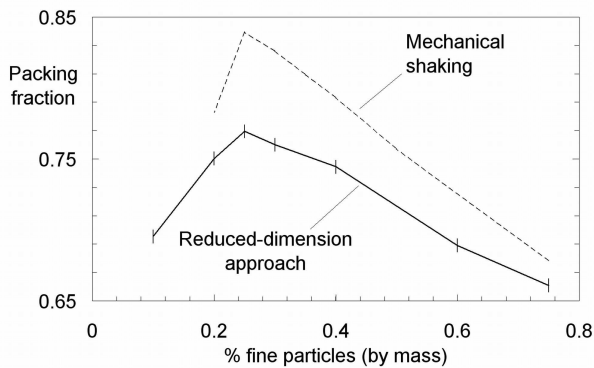}
\caption{\label{fig:comp16-1} Comparison of simulation results to
mechanical packing data \cite{mcgeary61} for a particle size ratio
of 248:15 (about 16.5:1).} \end{figure}

Investigations with binary packs of various size ratios show that
the difference between the measured packing fraction and that
produced by our simulation can be as large as several percent in
some cases (for large particle size differences); see Fig.
\ref{fig:comp16-1}. We suspect that the large difference arises
primarily because of the ballistic deposition method employed. The
ballistic deposition approach inevitably will produce lower packing
fractions than would be obtained with mechanical shaking, where the
particles may reorient themselves cooperatively as a group.

\begin{figure}
\includegraphics[scale=0.40]{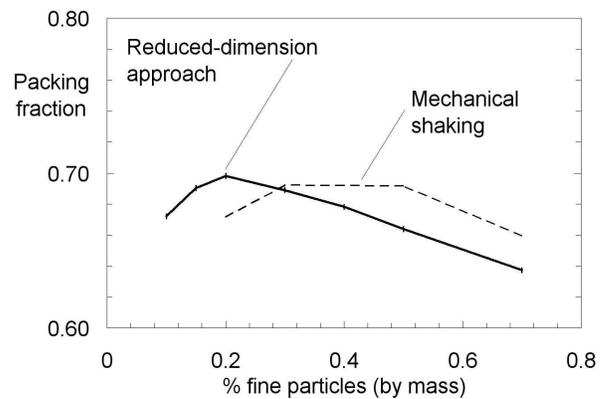}
\caption{\label{fig:comp3-1}Comparison of simulation results to
mechanical packing data \cite{mcgeary61} for a particle size ratio
of 31:9 (about 3.4:1).}
\end{figure}

Similar comparisons of the simulation to published data for smaller
size ratios yields a somewhat surprising result. We find a closer
quantitative agreement for a size ratio of 31:9 (about 3.4), but the
packing fraction peak occurs at a different value of the mass
fraction for our simulation than that published in \cite{mcgeary61}.
Fig. \ref{fig:comp3-1} shows a fairly broad peak near a mass
fraction of about 40 percent fine particles for the mechanical data,
but our simulation packing fractions are largest near a mass
fraction of 20 percent fine particles. We point out that this trend
in peak position is consistent with simulations at other particle
size ratios. We cannot determine conclusively why there is such a
distinct difference between the published data and our simulation at
small particle radius ratios. The data from \cite{mcgeary61} is
sparse in this region, and may not fully represent what might be
observed in a more exhaustive study.

We have investigated extensively the differences between packing
fraction results for reduced-dimension and three-dimensional
simulations for the data presented above. The data indicate a small
(about 0.5\%) bias between the reduced-dimension and
three-dimensional approaches, with an identical trend in shape. We
are confident that the larger differences between the data in
\cite{mcgeary61} and our simulations do not arise from the use of
the reduced-dimension approach, but rather are a feature of
ballistic deposition methods in general, whether the
three-dimensional or reduced-dimension method is used.

\subsection{Radial distribution functions}

In an effort to understand the microstructure of the packs produced
by our algorithm, we examined the radial distribution function of
the particles in both three-dimensional and reduced dimension packs.
Recent investigations of $G_{11}$ (the pair distribution function)
illustrate that it may be used as a tool to aid understanding of the
nature of order in a random particle pack, and includes features
that may be ascribed to jamming \cite{donev05}.

\begin{figure}
\includegraphics[scale=0.4]{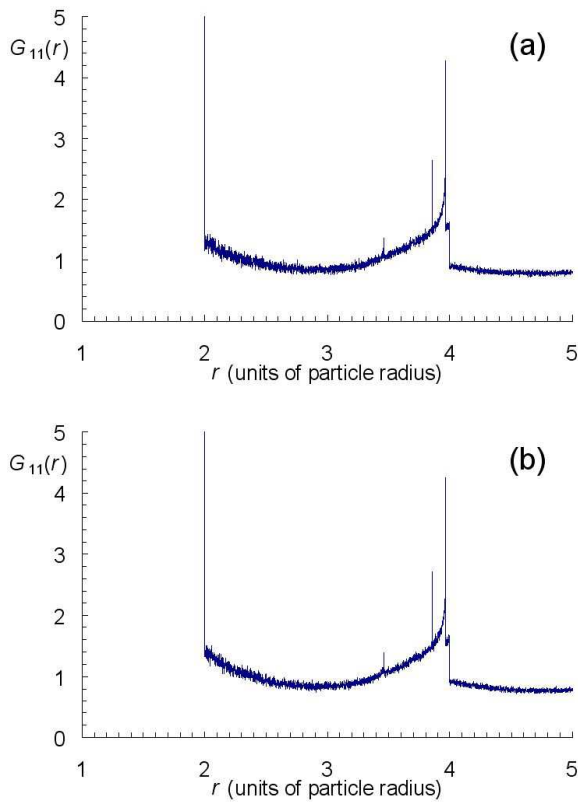}
\caption{\label{fig:3dcomp1}Pair distribution functions ($G_{11}$)
for a binary random pack (6.5:1 radius ratio) for a 3D simulation
(a) and a reduced-dimension simulation (b).}
\end{figure}

We first show results to support our claim that the
reduced-dimension approach produces identical radial distribution
functions to those obtained from 3D simulations. In figure
\ref{fig:3dcomp1}, we show the pair distribution function $G_{11}$
for a bimodal pack consisting of particles with a radius ratio of
6.5:1 for our simulation without any reduced-dimension algorithms
(inset a) and using the reduced-dimension algorithms and statistics
calculation methods (inset b). Although there are very minor
differences between the two data sets (due to statistical noise
resulting from the finite size of the pack), the data are
essentially indistinguishable. (See the text below for a discussion
of the pack structure giving rise to the observed resonances in the
distribution function.)

\begin{figure}
\includegraphics[scale=0.4]{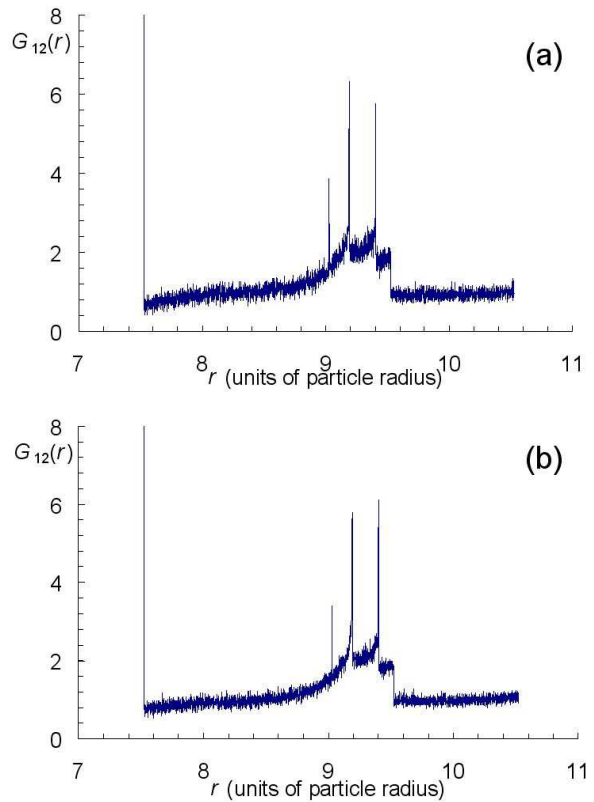}
\caption{\label{fig:3dcomp2}Radial distribution functions ($G_{12}$)
for a binary random pack (6.5:1 radius ratio) for a 3D simulation
(a) and a reduced-dimension simulation (b).}
\end{figure}

\begin{figure}
\includegraphics[scale=0.4]{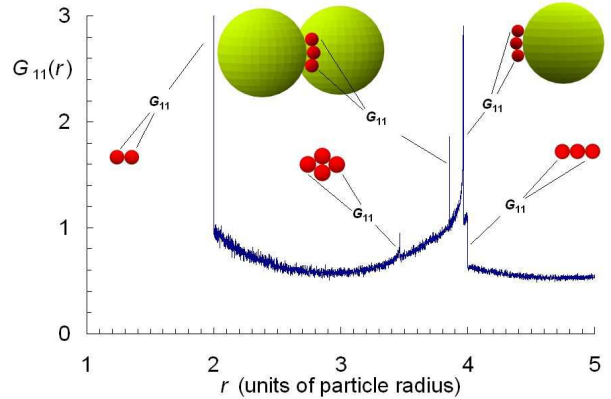}
\caption{\label{fig:g11-6.5to1}Radial distribution function $G_{11}$
for a binary pack consisting of two particles with radius ratio
6.5:1 (25\% fine by mass). An additional resonance at
$r\approx3.855$, not observed for smaller size ratios, appears due
to the line of three smaller particles wrapping around the roll
corridor formed by the two larger particles. This resonance exhibits
delta-function behavior because there is no freedom for the
particles to move slightly in their configuration.}
\end{figure}

\begin{figure}
\includegraphics[scale=0.4]{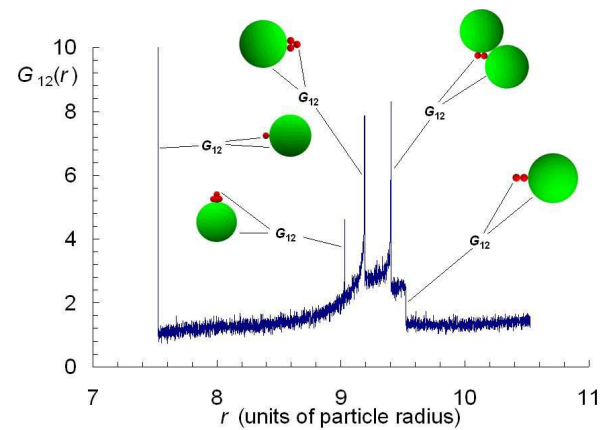}
\caption{\label{fig:g12-6.5to1}Radial distribution function $G_{12}$
for a binary pack consisting of two particles with radius ratio
6.5:1 (25\% fine by mass). The resonance at $r\approx9.029$ exhibits
delta-function behavior similar to that shown at $r\approx3.855$ in
Fig. \ref{fig:g11-6.5to1}.}
\end{figure}

We also show the radial distribution function $G_{12}$ for this same
binary pack in figure \ref{fig:3dcomp2}. Again, the two
distributions are essentially identical; the slight variation in
peak height for the resonances near $r\approx9$ is again due to the
finite size of the pack. (The data shown in figures
\ref{fig:3dcomp1} and \ref{fig:3dcomp2} is presented again in
figures \ref{fig:g11-6.5to1} and \ref{fig:g12-6.5to1} in the context
of the discussion of the pack microstructure.)

\begin{figure}
\includegraphics[scale=0.4]{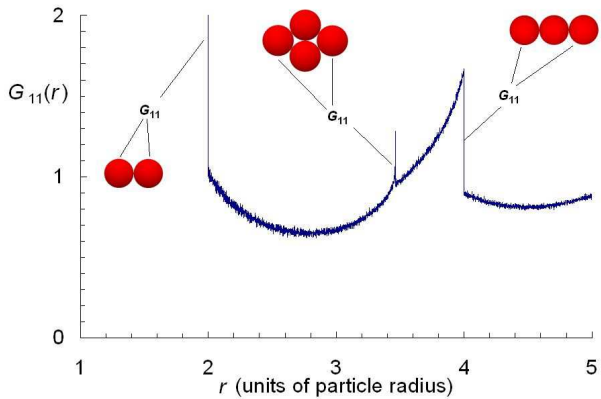}
\caption{\label{fig:g11smr-sph}Radial distribution function for a
monomodal random pack for small $r$. In this and the figures to
follow, the diagram shows the particle configurations required for
the resonance, and the labels indicate the particle pairs giving
rise to the resonance. In some cases, other particles may be
required to support the particles in their configuration, but these
particles are not shown.}
\end{figure}

Our radial distribution function for a monomodal pack follows
closely that of Donev, Torquato and Stillinger \cite{donev05}, with
two notable exceptions. First, we do not observe the $(r)^{-0.6}$
divergence near contact (where $r$ is the range away from contact).
We ascribe this to the nature of our simulation, which places
particles at a stability point in a static configuration, rather
than allowing them to reorient dynamically as in other simulations.
Second, we measure the ``second split peak'' at $\sqrt{3}D$ (where
$D$ is the particle radius), but the strength of this resonance is
much smaller in our packs than in the earlier work \cite{donev05};
in particular, the shoulders of the resonance are much less
pronounced; see Fig. \ref{fig:g11smr-sph}. The fact that this peak
is less pronounced in our pack suggests that the pack is less
ordered, and may be responsible in part for the known decrease in
packing density that exists for packs created with the ballistic
deposition method.

\begin{figure}
\includegraphics[scale=0.4]{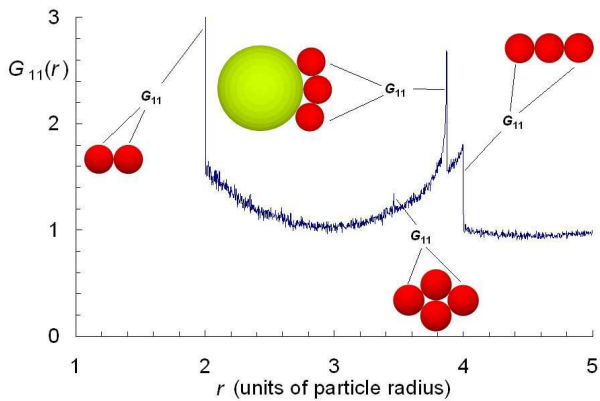}
\caption{\label{fig:g11-3to1}Radial distribution function $G_{11}$
for a binary pack consisting of two particles with radius ratio 3:1
(25\% fine by mass). The monomodal peak at $r=4$ splits into two
peaks when some of the smaller particles wrap around the larger
particle.}
\end{figure}

The pair distribution functions in binary packs exhibit rich
structure of a similar character to that shown in Fig.
\ref{fig:g11smr-sph}, but with more features. For example, the pair
distribution function $G_{11}$ for a binary pack with radius ratio
3:1 shows that the peak observed at $r=4$ in the monomodal pack
splits into two peaks. Some of the particles that give rise to this
feature in the monomodal pack (the ``line'' of three particles) are
wrapped around the surface of the larger particle in the binary
pack, so that the peak splits; see Fig. \ref{fig:g11-3to1}.

Increasing the size of the second particle in the binary pack above
4:1 shows that an additional resonance arises near the peak at
$r=4$. This additional resonance comes from a line of three smaller
particles that wrap around the roll corridor formed by two larger
particles. The roll corridor has a smaller radius than the radius of
the larger particle, yielding the satellite peak at smaller $r$.
This resonance does not occur in binary packs with a size ratio of
less than 4:1 because in that case, the smaller particles protrude
beyond the roll corridor formed by the larger particles, preventing
formation of this configuration. (See Fig. \ref{fig:g11-6.5to1}). We
note that this resonance (near $r\cong3.855$) is qualitatively
different from the others. In essence, the jammed state of this
particle configuration affords no room for the particles to shift
slightly from the default positions, so that the resonance at this
value of $r$ exhibits a delta-function behavior. The other
resonances have shoulders as $r$ approaches the resonance,
indicating slight variations in these configurations. Note that the
high resolution with which we measure the radial distribution
functions is a direct consequence of our reduced-dimension approach
and the capability to simulate systems with millions of particles
with large size differences.

Finally, we consider $G_{12}(r)$ for a binary mixture (mass ratio
25\% fine) with a radius ratio of 6.5:1. This function exhibits
resonances not seen, for example, in $G_{12}$ for a particle size
ratio of 3:1. In particular we find a delta-function resonance near
$r\approx9.029$. This resonance arises from a tetrahedron of small
particles whose base is in contact with the larger particle. Similar
to that shown in Fig. \ref{fig:g11-6.5to1}, this resonance
demonstrates no freedom of slight movement for the particle atop the
tetrahedron, in turn yielding a delta-function resonance at this
value of $r$; see Fig. \ref{fig:g12-6.5to1}. Smaller size ratios do
not exhibit this feature, presumably because the tetrahedron cannot
form on the surface unless the size ratio is large enough.

\subsection{A note about jamming}
The pair distribution functions produced from our simulation exhibit
features commonly associated with jamming \cite{donev05}. However,
the resonances found in the distribution functions simply arise from
ordered sets of particles in contact, and are not definitive proof
of jamming. Recent work has placed the understanding and
qualification of jamming on a rigorous foundation
\cite{donev04b,torquato01}.

At first blush, our simulation appears to produce hard-sphere
packings that cannot be jammed according to these recent
definitions. For example, the outer cylinders may contain particle
that ``hang'' in mid-air (i.e., supported by virtual particles).
However, we have verified by direct calculation that, in the
innermost cylinder (which by design is meant to replicate a ``3D''
packing), each particle is in fact locally jammed \cite{torquato01},
in that each particle has at least 4 contacts with other particles,
not all of which are in the same hemisphere. (As is common to other
numerically-simulated systems, our simulation produces a very small
number of ``rattlers'', or particles with only three contacts; the
typical abundance of such particles is about 0.1\% by number in a
monomondal, 3D simulation.) These observations suggest that further
work on the subject of jamming in a reduced-dimension simulation
might prove fruitful, and is one of our active lines of
investigation.

\section{Conclusion}

In summary, we have demonstrated a new, reduced-dimension, Monte
Carlo ballistic deposition algorithm that allows analysis of
multimodal hard-sphere systems. The simulation produces radial
distribution functions which were heretofore unavailable for
hard-sphere systems with large particle size differences. These
distribution functions exhibit resonances associated with particle
configurations found in the pack microstructure. This microstructure
information is important for researchers attempting to determine
macroscopic material properties using \textsl{ab initio}, first
principles calculations. These results also suggest that this
simulation technique may be a useful tool for further study of order
in hard-sphere systems with diverse particle sizes.

\section{acknowledgments} The authors would like to acknowledge and
thank Micheal Iverson of ATK Thiokol for sharing his expertise with
the design and development of the packing simulation and pack
viewing software.

\bibliography{text}

\end{document}